\documentclass{jaa}
\usepackage{natbib}
\usepackage{graphicx}
\usepackage{amsmath}	
\usepackage{amssymb}	
\usepackage{gensymb}
\usepackage{natbib}
\usepackage{xspace}
\usepackage{textcomp, gensymb}
\usepackage{longtable}
\usepackage{pdflscape}
\usepackage{float} 
\usepackage{footnote} 

\let\VANthebibliography\thebibliography
\def\thebibliography{\DeclareRobustCommand{\VAN}[3]{##3}\VANthebibliography}

\DeclareUnicodeCharacter{2212}{-}

\begin{document}\sloppy

\title{Estimation of Stellar Parameters and Mass Accretion Rate of Classical T Tauri Stars from LAMOST DR6\\ }


\author{S. Nidhi\textsuperscript{1}, Blesson Mathew\textsuperscript{1}, B. Shridharan\textsuperscript{1}, Suman Bhattacharyya\textsuperscript{1}, Edwin Das\textsuperscript{1} and Sreeja S Kartha\textsuperscript{1}}
\affilOne{\textsuperscript{1}Department of Physics and Electronics, CHRIST (Deemed to be University), Bangalore 560 029, India.\\}


\twocolumn[{

\maketitle

\corres{nidhi.sabu@res.christuniversity.in}

\msinfo{04 Nov 2022}{12 May 2023}{}
\begin{abstract}
Classical T Tauri stars are low-mass pre-main sequence stars with an active circumstellar environment. In this work we present the identification and study of 260 Classical T Tauri stars using LAMOST Data Release 6, among which 104 stars are newly identified. We distinguish Classical T Tauri stars from Giants and main-sequence dwarfs based on the log \textit{g} values and the presence of H$\alpha$ emission line and infrared excess that arises from the circumstellar accretion disk. We estimated the mass and age of 210 stars using the Gaia color-magnitude diagram. The age is from 0.1 to 20 Myr, where 90\% of the stars have age below 10 Myr and the mass ranges between 0.11 to 1.9 M\textsubscript{\(\odot\)}. From the measured H$\alpha$ equivalent widths, we homogeneously estimated the mass accretion rates for 172 stars, with most values ranging from $10^{-7}$ -- $10^{-10}$ $M_{\odot}yr^{-1}$. The mass accretion rates are found to follow a power law distribution with the mass of the star, having a relation of the form $\dot{M}_{acc}$ $\propto$ $M_{*} ^ {1.43 \pm 0.26}$,  in agreement with previous studies.

\end{abstract}

\keywords{emission-line star--Classical T Tauri star--accretion.}
}]



\doinum{12.3456/s78910-011-012-3}
\artcitid{\#\#\#\#}
\volnum{000}
\year{0000}
\pgrange{1--}
\setcounter{page}{1}
\lp{1}

\section{Introduction}

Gravitational collapse of cold interstellar clouds gives birth to the dense core in the protostellar phase. The dense core is embedded in the gas and dust associated with the molecular cloud. As it evolves, the embedded gas and dust will disappear, resulting in a circumstellar disk around it \citep{2008Sci...321..669Y, 2009ASPC..419..339S}. This stage of a star is called pre-main sequence (PMS) phase. PMS stars are young and are found in varying mass and size. Generally, they are classified as intermediate-mass and low-mass PMS stars. Young, early-type stars belonging to spectral types B, A, till mid-F-type, with a mass range of 2 -- 10 M\textsubscript{$\odot$} are intermediate-mass PMS stars, and those showing emission-lines in their spectrum are Herbig Ae/Be stars (HAeBe; \citealp{1960ApJS....4..337H}). The low-mass ($<$ 2 M\textsubscript{$\odot$}) counterparts are T Tauri stars (TTS) which were defined by Joy (1945) as objects associated with nebulosity showing characteristic emission-line spectrum. These stars belong to the spectral type ranging from mid-F-type till late M-type and are found to be systematically brighter than main-sequence stars of the same spectral type \citep{1997AJ....113.1733H}. 

Generally, TTS are classified as classical-T Tauri stars (CTTS) and weak-line T Tauri stars (WTTS). CTTS have strong magnetic fields \citep{1999A&A...341..768G, 2003ApJ...584..911F} through which matter is channeled from the inner circumstellar disk to the stellar surface. The hotspots formed due to the infalling material on the stellar surface results in excess emission in ultraviolet and optical continuum \citep{1998ApJ...492..323G, 1991ApJ...370L..39K, 2016ARA&A..54..135H}. Along with these signatures, CTTS show line emission (particularly H$\alpha$), and infrared (IR) excess due to the presence of circumstellar disk \citep{1998ApJ...492..323G, 2003ApJ...592..266M}. On the contrary, WTTS present little or no evidence for accretion and IR excess as compared to CTTS \citep{1995ApJ...452..736H}. Henceforth, WTTS can be assumed to be in a more advanced stage of stellar evolution, with no circumstellar accretion disks, indicating the origin of H$\alpha$ emission line entirely from chromospheric activities \citep{1989ARA&A..27..351B, 1994ApJ...426..669H}. Thus, several studies have proposed IR excess as well as the equivalent width of H$\alpha$ [EW(H$\alpha$)] emission line as criteria to distinguish CTTS from WTTS. In addition, homogeneous spectroscopic studies of these young stellar objects (YSO) opens up the possibility of identifying CTTS and estimating their mass accretion rates \citep{2003ApJ...592..266M, 2006A&A...452..245N, 2008ApJ...681..594H, 2015MNRAS.453.1026K, 2017A&A...604A.127M}. 

Generally, the well-studied CTTS from the literature are located around the solar neighborhood or towards the Galactic center \citep{2007ApJ...667..308C, 2011MNRAS.415..103B, 2011ApJS..195....3F, 2012A&A...548A..56R, 2014ApJ...786...97H, 2019MNRAS.484.5102K}. In this work, we made use of the data release from the Large Sky Area Multi-Object Fiber Spectroscopic Telescope (LAMOST), to identify young PMS population in the Galactic anti-center direction. Pioneering work in the search of early-type PMS stars using the LAMOST data towards this region were carried out by \cite{2016RAA....16..138H}, \cite{2021MNRAS.501.5927A}, \cite{2021RAA....21..288S}, \cite{2022ApJ...936..151Z} and Nidhi et al. (2022; under review). These studies have significantly increased the sample of known early-type PMS stars in the Galaxy. On the other hand, very few studies on CTTS is carried out in this region, such as that of \cite{2018Ap&SS.363..104L}, who found 38 CTTS with X-ray detection and H$\alpha$ emission line. Also, \cite{2022ApJ...936..151Z} identified 20 F-type PMS stars, of which 4 are T Tauri stars. Hence, in this work, we are homogeneously analyzing the CTTS towards the Galactic anti-center region and evaluating the stellar and accretion properties. In section 2 we discuss the LAMOST survey program and the sample selection method employed to identify the CTTS stars from the LAMOST catalog. Section 3 describes the spatial distribution, evolutionary status and accretion properties of CTTS stars from this study. The results are summarized in Section 4.

\section{Data Analysis}

In this section, we provide a brief overview of the LAMOST telescope and about the observational strategy that provided the data for this study. Also, we explain the step-by-step criteria used to identify CTTS from LAMOST catalog.

\subsection{Defining the sample set from the LAMOST data release}

The Large sky Area Multi-Object fiber Spectroscopic Telescope (LAMOST) is a 4 m quasi-meridian reflecting Schmidt telescope, with a field of view of 20 $deg^{2}$ \citep{2012RAA....12.1197C, 2012arXiv1206.3569Z}. It is also known as Guo Shoujing Telescope, maintained by
Xinglong station of the National Astronomical Observatories, Chinese Academy of Sciences. LAMOST is equipped with 4000 optical fibers in its focal plane and can obtain up to 4000 spectra in one exposure \citep{2012arXiv1206.3569Z}. The program was initiated in 2012 September and completed the pilot survey and five-years of phase {\sc I} survey. Phase {\sc II} of the survey started from 2018 with the 6th data release (DR6{\footnote{https://dr6.lamost.org}}), containing 9,911,337 low-resolution spectra (LRS). The LRS has a resolution of R$\sim$1800 at 5500 \r{A}~and wavelength coverage of 3700 $\r{A} \leq \lambda \leq 9000 \r{A}$. All of the LRS were generally classified into different categories through the pipeline, including 9,231,057 stellar spectra, 177,270 galaxy spectra, 62,168 quasar spectra, and 440,842 spectra marked as unknown. We retrieved the spectra of 8,613,842 late-type stars belonging to the spectral type of F5-M9 from the LAMOST DR6 catalog, and is used for our analysis.

\subsection{Procedure for the identification of Classical T-Tauri stars}

\subsubsection{Finding late-type emission line stars}

H$\alpha$ emission line is a prominent feature observed in the emission-line stars (ELS; \citealp{2007ASSL..342.....K}). An automated python routine was used, that employs \texttt{find{\textunderscore}peaks} function from the \texttt{scipy} package to detect the H$\alpha$ peak within a wavelength window of 6561--6568 \AA. Based on the presence of the H$\alpha$ emission line, Edwin et al. 2022 (in prep) identified 77,586 late-type ELS (LELS) belonging to the spectral type of F5-M9 from LAMOST DR6.  We used the 77,586 LELS as our initial sample, in search of CTTS from LAMOST DR6. It was noted that some sources have multiple observations from LAMOST. Hence, we retained the spectrum with Signal-to-Noise Ratio in SDSS r-band value greater than 10 ($SNR_{r}$ $>$ 10; \citealp{2016RAA....16..138H}). This reduced the number of LELS to 56,576. The photometric magnitudes from Gaia DR3 \citep{2022yCat.1355....0G} mission and its distance estimates from \cite{2021AJ....161..147B} are queried for the LELS, resulting in 49,392 stars.

\begin{figure}[H]   
    \includegraphics[width=\columnwidth]{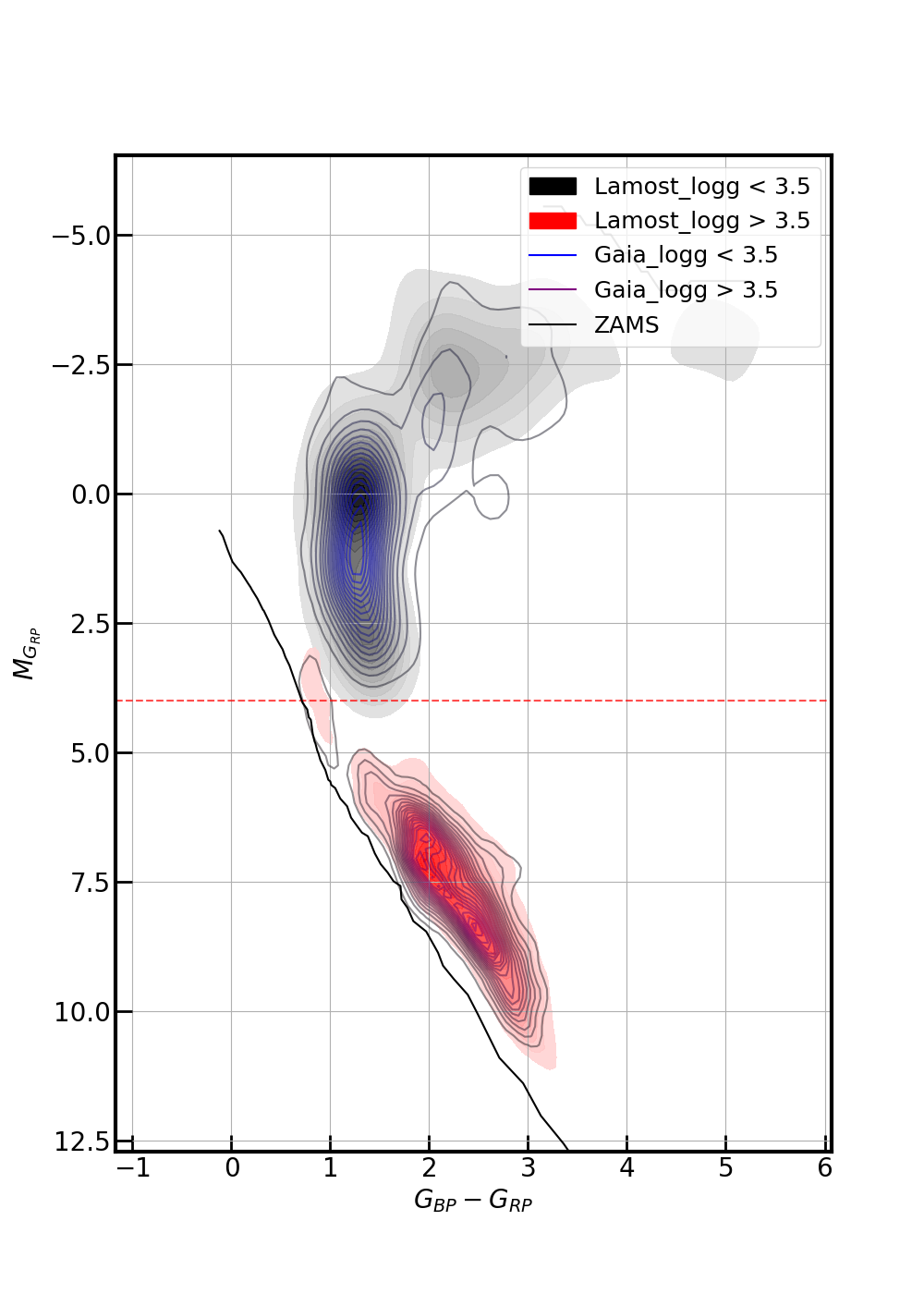}
    \caption{Probability distribution (Gaussian fitted at 20 contour levels) of LELS in the \textit{Gaia} CMD is represented in the figure. The log \textit{g} values of LELS from the LAMOST DR6 are shown in black and red shaded colors, respectively. The black shade denotes stars with log \textit{g} $<$ 3.5, while the red shade denotes stars with log \textit{g} $>$ 3.5. Also, the log \textit{g} values of LELS from the Gaia DR3 catalog are displayed as blue and violet line contours, for evolved stars and main sequence stars, respectively. The contours of the blue indicate log \textit{g} $<$ 3.5, and violet lines indicate log \textit{g} $>$ 3.5 The red dashed line indicates the $M_{G_{RP}}$ value of 4.0. The black line represents ZAMS from \cite{2013ApJS..208....9P}.}
    \label{fig:CMD_log}
\end{figure}

Among 49,392 sources, we noted that 50\% of the stars belongs to evolved star category (giants/supergiants). These are highly luminous, post-main sequence stars reported to have H$\alpha$ emission \citep{1973ApJ...186..909R}. The H$\alpha$ emission in evolved stars primarily originates from wind-driven mass loss \citep{1963ARA&A...1...97W, 1970MNRAS.147..161H}. Several studies in the literature have categorized evolved stars based on the surface gravity (log \textit{g}), as for example, see the studies of \cite{2016MNRAS.461.3336C} (log \textit{g} $<$ 3), \cite{2021MNRAS.505.5340M} (3.2 $\geq$ log \textit{g} $\geq$ -1.0) and \cite{2022A&A...664A..78F} (log \textit{g} $<$ 3.5). In the present work, to separate evolved stars from LELS, we set a threshold of log \textit{g} = 3.5, as shown in Figure \ref{fig:CMD_log}. We used the log \textit{g} values provided by LAMOST DR6, which was estimated using LAMOST stellar parameter pipeline (LASP; \citealp{2011RAA....11..924W, 2015RAA....15.1095L}). Interestingly, log \textit{g} estimates are available from Gaia DR3 for the sample of stars used in this study. Hence, we were able to assess the reliability of log \textit{g} values from LAMOST DR6 by comparing it with Gaia DR3 values (see Figure \ref{fig:CMD_log}).

In Figure \ref{fig:CMD_log}, we have presented the probability distributions (Gaussian fitted at 20 contour levels) of the sample of stars in the Gaia color-magnitude diagram (CMD), in a manner similar to the representation in \cite{2022ApJ...933L..34B}. In the CMD plot, we used the Gaia broad-band photometric magnitudes, \textit{G\textsubscript{BP}}, and \textit{G\textsubscript{RP}} from Gaia DR3 to construct the Gaia CMD. \textit{G\textsubscript{BP}} and \textit{G\textsubscript{RP}} are corrected for extinction using the 3D dust map of \cite{2019ApJ...887...93G} and distance estimates from \cite{2021AJ....161..147B}. From the figure, the distribution makes it evident that the entire LAMOST primary sample can be divided into two parts near $M_{G_{RP}}$ = 4.0. The giant branch is surrounded by stars with log \textit{g} $<$ 3.5, while stars located near to ZAMS are populated with log \textit{g} $>$ 3.5. Gaia DR3 sample serves as a point of reference for similar distribution, which enables us to confirm the LAMOST data distributions. Since the present work is concerned with CTTS, we removed the stars with log \textit{g} $<$ 3.5. Henceforth, we consider 24,745 LELS for further analysis.

\subsubsection{Selection of CTTS using 2MASS--WISE color--color diagram}

In CTTS, the IR excess is commonly attributed to the presence of dust in their inner circumstellar disk, which decreases over time as the disk material is dissipated \citep{1979ApJS...41..743C, 2006ApJ...648..484H}. Hence, to extract CTTS from LELS, we used the Two Micron All Sky Survey (2MASS; \citealp{2003yCat.2246....0C}) and Wide-field Infrared Survey Explorer (WISE; \citealp{2014yCat.2328....0C}) IR photometric data to construct a color-color diagram (CCDm). The AllWISE program extends the work of the successful WISE mission by producing a new source catalog and image atlas with enhanced sensitivity and accuracy. However, the ALLWISE point source catalog is affected by high numbers of spurious sources, especially in the longer-wavelength bands. \cite{2014ApJ...791..131K} recommended a procedure to improve the source reliability in the WISE bands using signal to noise ratio (S/N) and reduced chi-squared values (${\chi^{2}_{\nu}}$). The criteria we used in this work adopted from \cite{2014ApJ...791..131K} to suppress false source contamination in WISE \textit{W1} and \textit{W2} bands are as follows.\newline
\\
WISE band 1: non-null \textit{w1sigmpro} and \textit{w1rchi2} $<$ (\textit{w1snr} - 3)/7. \newline
WISE band 2: non-null \textit{w2sigmpro}.\newline

\begin{figure}
    \includegraphics[width=\columnwidth]{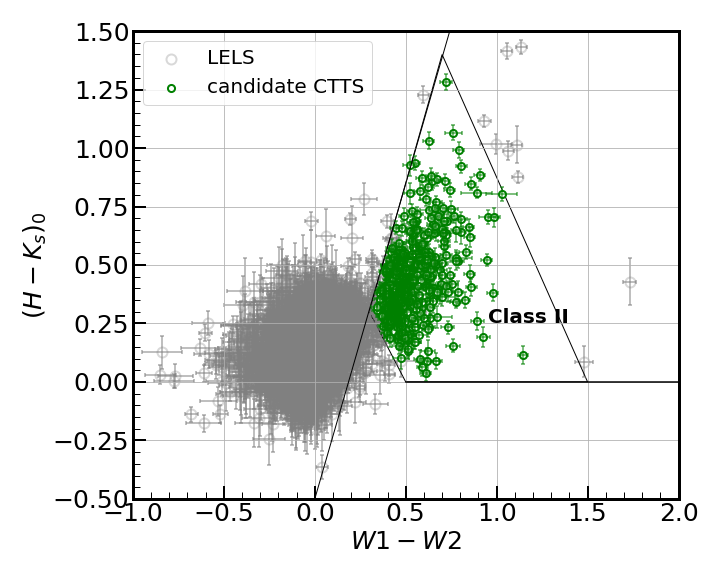}
    \caption{2MASS-WISE CCD. Gray points represent the LELS point sources. Green points within the polygon represent 374 candidate CTTS belonging to Class {\sc II} category showing IR excess. The YSO selection criteria is adopted from \cite{2014ApJ...791..131K}. }
    \label{fig:2WCCD}
\end{figure}
The \textit{w1sigmpro} and \textit{w2sigmpro} are the errors in \textit{W1} and \textit{W2} magnitude, respectively. \textit{w1snr} is the signal to noise ratio for W1 filter and \textit{w1rchi2} refers to the reduced chi-squared values of the W2 profile fit, as provided in the catalog. We also queried \textit{J}, \textit{H}, and \textit{$K_{s}$} magnitudes from the 2MASS point source catalog \citep{2003yCat.2246....0C}. The extinction parameter, $A_{V}$, values were retrieved from 3D dust map of \cite{2019ApJ...887...93G}, which were used for further analysis. The reddening corrected 2MASS-WISE CCDm of the LELS is shown in Figure \ref{fig:2WCCD}. The location of the search box for Class {\sc II} sources in Figure \ref{fig:2WCCD} is generated using the YSO search criteria given by \cite{2014ApJ...791..131K}. Class {\sc II} is a YSO classification category based on the continuum slope in the IR region of the spectral energy distribution (SED) of the star \citep{1987IAUS..115....1L, 1993ApJ...406..122A}. This phase follows that of a protostar, where the star has almost entirely dispersed its envelope but is still actively accreting from the optically thick accretion disk. We found 374 Class {\sc II} sources showing IR excess and hence can be considered as candidate CTTS. 

\subsubsection{H$\alpha$ Equivalent Width as an Empirical Criterion to identify CTTS}

\begin{figure}
    \includegraphics[width=\columnwidth]{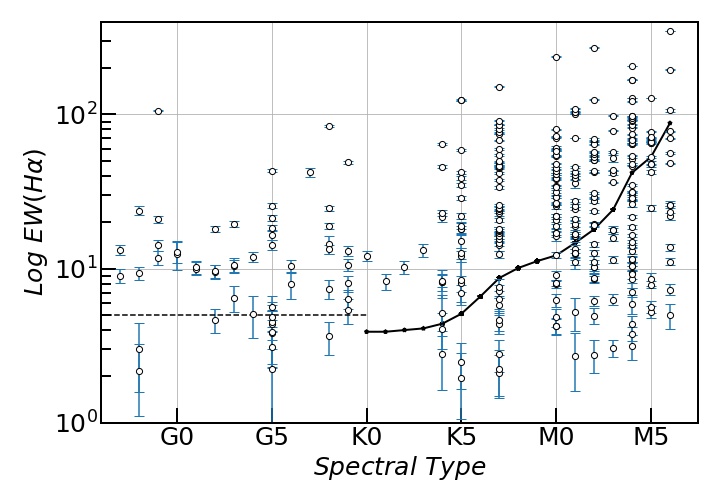}
    \caption{H$\alpha$ EW for 374 candidate CTTS is plotted against the spectral type. The thick curve represents the \cite{2003AJ....126.2997B} to distinguish CTTS from WTTS, for the spectral type from K0 to M7. For stars belonging to F5 to G9 spectral type, a minimum EW(H$\alpha$) of 5 \AA~is represented in a dotted line.}
    \label{fig:EW_cut}
\end{figure}

For CTTS, H$\alpha$ is due to the recombination radiation from the accretion disk. Hence, the emission strength of H$\alpha$ line is one of the simplest criteria to differentiate the CTTS from WTTS. \cite{1998AJ....115..351M} proposed the following dependence on the H$\alpha$ equivalent width [EW(H$\alpha$)] of CTTS with respect to the spectral type: 5 \AA~ for spectral types earlier than M0, 10 \AA~for M0–M2, and 20 \AA~for later types. \cite{2003ApJ...582.1109W} also empirically suggested the dependency of EW(H$\alpha$) with spectral type. Later, \cite{2003AJ....126.2997B} proposed EW(H$\alpha$) values as a function of spectral type, derived from the observed saturation limit for the chromospheric activity at Log($L_{H\mathrm{\alpha}}$/$L_{bol}$) = −3.3. Adopting the criterion given by \citealp{2003AJ....126.2997B}, (as shown in Figure \ref{fig:EW_cut}), we found 213 CTTS stars located above the solid curve. The EW(H$\alpha$) H$\alpha$ is measured using Image Reduction and Analysis Facility (IRAF; \citealp{1986SPIE..627..733T}) software. It should be noted that only stars belonging to the spectral type ranging from K0 to late M-type can be classified using the above criterion. For rest of the 47 other stars with spectral types from F5 to G9, we put forth a minimum EW(H$\alpha$) value of 5 \AA~based on the criteria given by \cite{1998AJ....115..351M}. Thus, by combining the H$\alpha$ emission strength and dust emission properties from the accretion disk, we obtained a sample of 260 CTTS whose spectral properties can be explored further.

\section{Results and Discussion}

The spatial location and the stellar properties of 260 CTTS identified from this work is analyzed in this section. We cross-matched our stars with the TTS/YSO catalogs {\footnote{The well-known TTS/YSO catalogs can be retrieved from the studies of \cite{1999AJ....118.1043H}, \cite{2009ApJS..181..321E}, \cite{2009ApJS..184...18G}, \cite{2011ApJS..195....3F}, \cite{2013ApJS..205....5H}, \cite{2014ApJ...786...97H}, \cite{2018ApJS..236...27C}, \cite{2019AJ....157...85B}, \cite{2005ApJ...624..808L}, \cite{2013ApJ...764..172P}, \cite{2014ApJ...788..102Z}, and \cite{2020A&A...640A.127Z}}} as well as with the SIMBAD database and found identifiers for 156 stars, in which 86 stars are labeled as TTS, 6 as TTS candidate, 42 as YSO, 12 as YSO candidate and 10 as ‘stars’. Though 49 stars from the remaining sample of 104  were mentioned in the study of \cite{2020A&A...643A.122S}, they haven't classified them into any category of ELS. Hence, through this work, we have identified a new sample of 104 CTTS for which its spatial distribution, stellar and accretion properties are evaluated.

\subsection{Spatial Distribution of CTTS in the Galactic plane}

\begin{figure}[h]
    \includegraphics[width=\columnwidth]{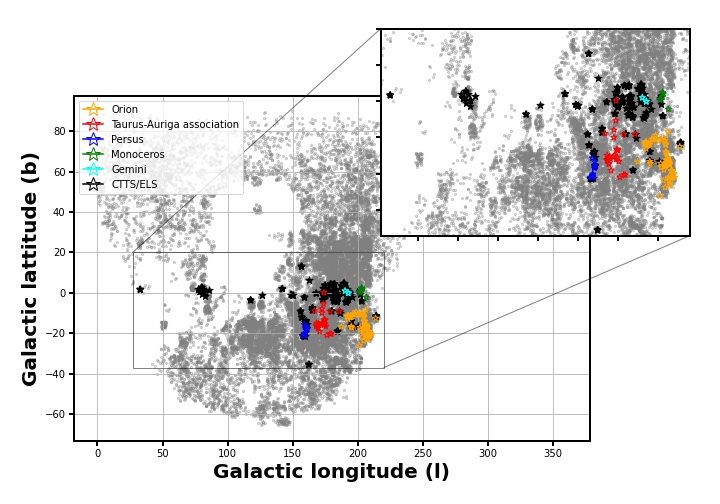}
    \caption{The plot represents the spatial distribution of 260 CTTS stars in Galactic Latitude (b) vs Galactic Longitude (l) plane. It can be seen that most of the ELS are found to be within $100\degree \leq l \leq 220\degree$ and latitudes $-35\degree \leq |b| \leq15\degree$. The grey points represents the 24,745 LELS. Stars in orange, red, blue, green and  cyan color belong to Orion, Taurus-Auriga association, Perseus, Monoceros and Gemini star forming regions, respectively. Stars represented in black color are not identified to be part of any star-forming regions.}
    \label{fig:space}
\end{figure}

LAMOST Spectroscopic Survey of the Galactic anti-center (LSS—GAC) has provided a unique opportunity to explore the distribution of CTTS along the Galactic anti-center direction \citep{2014IAUS..298..310L}. The spatial distribution of CTTS is  represented in the Galactic coordinates in Figure \ref{fig:space}. It is observed that 90\% of CTTS are spread over the Galactic longitudes $100\degree \leq l \leq 220\degree$ and latitudes $-35\degree  \leq |b| \leq 15\degree$, implying that the CTTS from this study are spread more towards the Galactic anti-center direction. In Figure \ref{fig:space}, we have represented 142 CTTS based on the star forming regions with which they are associated. Also, 63 stars are classified as `ELS' but they are not reported to be associated with any known star forming regions. The remaining  55 stars have not been reported in any catalogs and can be considered as new detections. Further investigation on the membership analysis of these stars will be carried out as future work.

\subsection{Stellar properties}

\begin{figure}
    \includegraphics[width=\columnwidth]{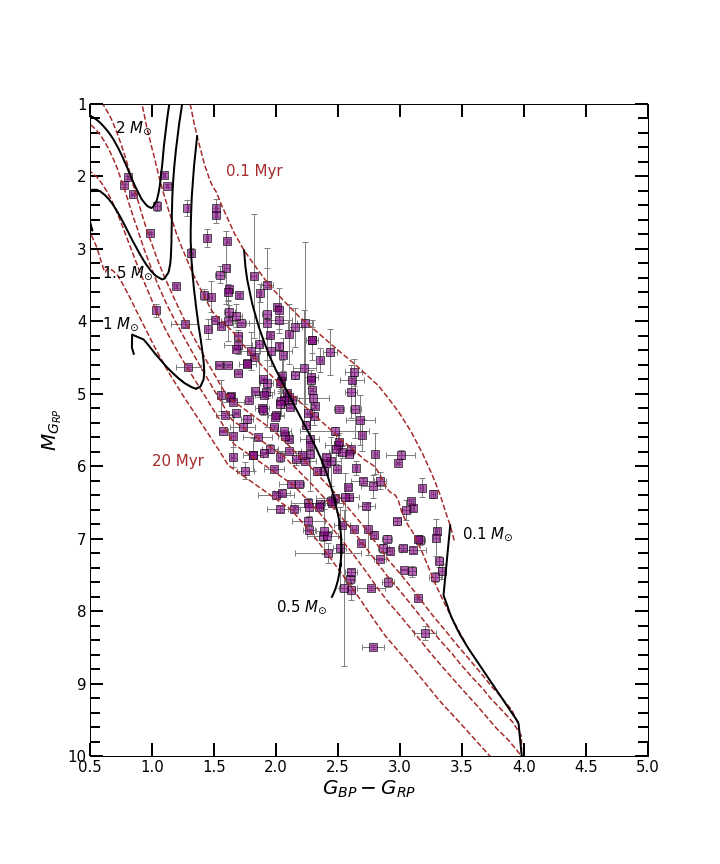}
    \caption{Gaia CMD for CTTS. The isochrones (dashed lines) and tracks (solid lines) are from \cite{2012MNRAS.427..127B} at solar metallicity.}
    \label{fig:age_mass}
\end{figure}

The stellar age ($t_{*}$) and mass ($M_{*}$) of CTTS are estimated by placing them in the extinction corrected (see sect. 2.2.1) \textit{Gaia} CMD and over-plotted with the isochrones and theoretical evolutionary tracks, respectively (Figure \ref{fig:age_mass}). We employ \cite{2012MNRAS.427..127B} solar metallicity PMS isochrones and tracks to estimate the age and mass for the CTTS. 

We obtained the age for 210 CTTS from the \textit{Gaia} CMD in Figure \ref{fig:age_mass}. The age range of CTTS derived from \cite{2012MNRAS.427..127B} isochrones is from 0.1 -- 20 Myr, where 90\% of the stars have age below 10Myr. Similarly, from Figure \ref{fig:age_mass}, the mass for 210 CTTS lying above the ZAMS line is estimated from the \textit{Gaia} CMD, over-plotted with the \cite{2012MNRAS.427..127B} evolutionary tracks. The estimated mass range of the CTTS is between 0.11 to 1.9 M\textsubscript{\(\odot\)}. The extinction, distance and the estimated age and mass are given Table \ref{tab:T1}.

\subsection{Accretion properties}

CTTS have a circumstellar disk from which the matter is accreted onto the stars \citep{1989A&ARv...1..291A}. In this scenario, the central star has a strong magnetic field producing large starspots. The magnetic field lines truncate the circumstellar disk at a few radii from the star, also known as the truncation radius ($R_{in}$). The matter from the disk gets accreted onto the star through the magnetic field lines, forming accretion columns or funnel flows. The infalling gas moves under free-fall velocities of the order 300 $km/s$, causing shocks near the stellar surface. The resulting energy generated from magnetospheric accretion, heats and ionizes the circumstellar gas, thereby causing emission lines \citep{2016ARA&A..54..135H}. 

The re-radiated energy can be measured from the emission line luminosity which is correlated to the accretion luminosity ($L_{acc}$; \citealp{2017A&A...600A..20A}). In accreting stars, it has been shown that the major component of the H$\alpha$ emission arises from the accretion funnels and overlaps the emission produced by the accretion shock over the stellar photosphere \citep{2017A&A...600A..20A, 2017A&A...602A..33F}. We, therefore, determined the H$\alpha$ line luminosity ($L_{H\mathrm{\alpha}}$) from the EW(H$\alpha$) to estimate $L_{acc}$. To calculate $L_{H\mathrm{\alpha}}$, we first estimate the H$\alpha$ line flux $(F_{H\mathrm{\alpha}})$ using,
\begin{equation}
    F_{H\mathrm{\alpha}} = W(H\mathrm{\alpha}) \times  F_{R}(H\mathrm{\alpha})   
\end{equation}
where $F_{R}(H\mathrm{\alpha})$ is the continuum flux density at H$\alpha$. The $F_{R}(H\mathrm{\alpha})$ is calculated using the extinction corrected \textit{R} band magnitude ($R_0$, \citealp{2018ApJ...857...30M}). Further, the $F_{H\mathrm{\alpha}}$ is converted to line luminosity using the relationship,
\begin{equation}
    L_{H\mathrm{\alpha}}=4\pi d^2 F_{H\mathrm{\alpha}}    
\end{equation}
where d is the distance to the star compiled from \cite{2021AJ....161..147B}. We used the following relation given in \cite{2003ApJ...592..266M} and \cite{2011MNRAS.415..103B} to calculate $L_{acc}$.
\begin{equation}
   Log({L_{acc}}) = 1.13(\; \pm \;0.07) \times Log({L_{H\mathrm{\alpha}}}) + 1.93(\;\pm\;0.23) 
\end{equation}
The $\dot{M}_{acc}$ can then be calculated from the free-fall equation:
\begin{equation}
L_{acc} = \frac{GM_*\dot{M}_{acc}}{R_*}(1-\frac{R_*}{R_{in}})
\end{equation}
where $M_*$ and $R_*$ are the stellar mass and radius, respectively. Based on the previous studies \protect\citep{1998ApJ...492..323G, 2008ApJ...681..594H, 2012A&A...548A..56R, 2014A&A...561A...2A, 2016A&A...591L...3M}, we adopt $R_{in} = 5R_{*}$.

\begin{figure}
    \includegraphics[width=\columnwidth]{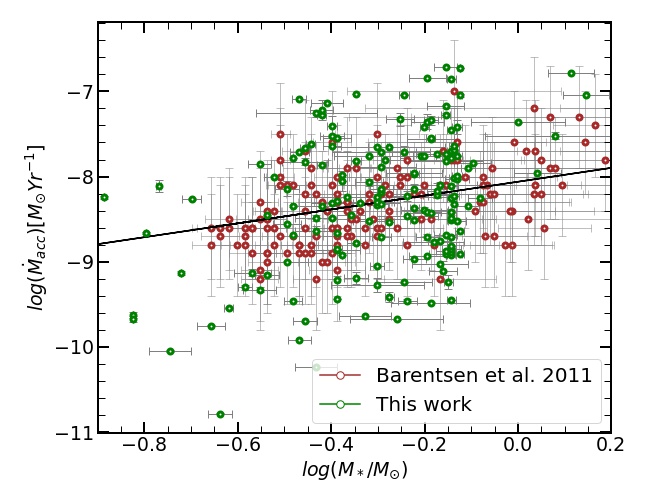}
    \caption{The figure shows the log-log plot between $\dot{M}_{acc}$ -- $M_*$ relation of CTTS. Stars from this work are given in green and the sample from \cite{2011MNRAS.415..103B} is given in brown color. The solid black line represents the corresponding best fit to the data with a slope of 0.81 $\pm$ 0.17.  }
    \label{fig:Macc}
\end{figure}

The $\dot{M}_{acc}$ is estimated for 172 CTTS, with most values ranging from $10^{-7}$ -- $10^{-10}$ $M_{\odot}yr^{-1}$. The $\dot{M}_{acc}$ of CTTS correlates with the stellar mass, following the power-law relation of $\dot{M}_{acc}$ $\propto$ $M_{*} ^ {\alpha}$. In this study, we obtained the power-law of $\dot{M}_{acc}$ $\propto$ $M_{*} ^ {1.43 \pm 0.26}$ as the best fit to the data from this work, which is in agreement with $\dot{M}_{acc}$ $\propto$ $M_{*} ^ {1.1 \pm 0.2}$ obtained by \cite{2011MNRAS.415..103B}. In figure \ref{fig:Macc}, we compared the sample with the \cite{2011MNRAS.415..103B} sample, where the stars are found to follow the similar trend with a slope of 0.81 $\pm$ 0.17, for the combined sample. Whereas, previous studies have presented higher correlation of accretion rates with stellar mass, with $\alpha$ = 1.8 - 2.2 \citep{2003ApJ...592..266M, 2006A&A...452..245N, 2008ApJ...681..594H, 2015MNRAS.453.1026K}. \cite{2009A&A...504..461F} found $\alpha$ = 3.1, which is much steeper than the literature values. The difference in the values
is primarily due to the bias in the primary sample. In future work, we aim to conduct a more detailed investigation into the relationship between the astrometric and spectroscopic parameters in each star-forming region and compare the results. 

\begin{table*}[htbp]
    \caption{Parameters for five stars are given below. Spectral type from LAMOST, extinction values retrieved from \citet{2019ApJ...887...93G}, distance estimates of these stars obtained from \citet{2021AJ....161..147B}, estimated age, mass, and mass accretion rate for the individual stars from this study are provided. Full version of this table will be made available online. }
    \begin{tabular}{cccccccc}
    \label{tab:T1}\\
    \hline\hline
        \multicolumn{8}{c}{} \\
        LAMOST\_ID & Spectral type & EW(H$\alpha$) & A\textsubscript{V} & Distance & Age & Mass & log($\dot{M}_{acc}$) \\
         &  & (\AA) & (mag) & (pc) & (Myr) &(M\textsubscript{\(\odot\)}) & ($M\textsubscript{\(\odot\)}yr^{-1}$) \\
        \hline\hline

        J060413.19+211639.6	&	F7	&	-13.2$\pm$1	&	2.3542	&	$3636.7${\raisebox{0.5ex}{\tiny$^{+4803.5}_{-2683.9}$}}	&	$0.35${\raisebox{0.5ex}{\tiny$^{+0.8}_{-0.1}$}}	&	$0.51${\raisebox{0.5ex}{\tiny$^{+0.63}_{-0.42}$}}	&	$-7.70${\raisebox{0.5ex}{\tiny$^{+0.58}_{-0.58}$}}	\\
        J052911.44-060805.4	&	F7	&	-8.99$\pm$0.94	&	0.2194	&	$352.80${\raisebox{0.5ex}{\tiny$^{+355.90}_{-349.49}$}}	&	$4.5${\raisebox{0.5ex}{\tiny$^{+4.8}_{-4.3}$}}	&	$1.9${\raisebox{0.5ex}{\tiny$^{+1.9}_{-1.9}$}}	&	$-6.52${\raisebox{0.5ex}{\tiny$^{+1.17}_{-1.08}$}}	\\
        J061509.10+213319.8	&	F8	&	-23.8$\pm$1.6	&	1.5334	&	$1803.0${\raisebox{0.5ex}{\tiny$^{+2000.5}_{-1640.6}$}}	&	$0.6${\raisebox{0.5ex}{\tiny$^{+0.8}_{-0.5}$}}	&	$0.48${\raisebox{0.5ex}{\tiny$^{+0.55}_{-0.4}$}}	&	$-7.76${\raisebox{0.5ex}{\tiny$^{+0.55}_{-0.54}$}}	\\
        J051645.56-015122.3	&	F8	&	-9.33$\pm$0.93	&	0.4113	&	$367.86${\raisebox{0.5ex}{\tiny$^{+369.71}_{-365.10}$}}	&	$4.6${\raisebox{0.5ex}{\tiny$^{+4.9}_{-4.3}$}}	&	$1.85${\raisebox{0.5ex}{\tiny$^{+1.85}_{-1.85}$}}	&	$-7.63${\raisebox{0.5ex}{\tiny$^{+0.94}_{-1.01}$}}	\\
        J061141.85+203158.9	&	F9	&	-11.7$\pm$1.2	&	1.7208	&	$1813.9${\raisebox{0.5ex}{\tiny$^{+2106.2}_{-1572.5}$}}	&	$0.1${\raisebox{0.5ex}{\tiny$^{+0.3}_{-0.1}$}}	&	$0.32${\raisebox{0.5ex}{\tiny$^{+0.34}_{-0.29}$}}	&	$-8.14${\raisebox{0.5ex}{\tiny$^{+0.36}_{-0.53}$}}	\\
        J061544.65+240050.5	&	F9	&	-14.2$\pm$1.1	&	1.0695	&	$1920.3${\raisebox{0.5ex}{\tiny$^{+3072.0}_{-1238.3}$}}	&	$0.4${\raisebox{0.5ex}{\tiny$^{+2}_{-0.1}$}}	&	$0.46${\raisebox{0.5ex}{\tiny$^{+0.55}_{-0.38}$}}	&	$-8.31${\raisebox{0.5ex}{\tiny$^{+0.34}_{-0.33}$}}	\\
        J061246.81-061120.9	&	F9	&	-20.9$\pm$1.1	&	0.5126	&	$736.82${\raisebox{0.5ex}{\tiny$^{+776.73}_{-695.37}$}}	&	$0.45${\raisebox{0.5ex}{\tiny$^{+0.5}_{-0.4}$}}	&	$0.7${\raisebox{0.5ex}{\tiny$^{+0.72}_{-0.56}$}}	&	$-7.81${\raisebox{0.5ex}{\tiny$^{+0.58}_{-0.58}$}}	\\
        J055532.35+274451.3	&	F9	&	-106$\pm$0	&	1.1516	&	$1217.9${\raisebox{0.5ex}{\tiny$^{+1354.0}_{-1098.0}$}}	&	$9.3${\raisebox{0.5ex}{\tiny$^{+9.8}_{-6.4}$}}	&	$0.73${\raisebox{0.5ex}{\tiny$^{+0.73}_{-0.72}$}}	&	$-8.32${\raisebox{0.5ex}{\tiny$^{+0.37}_{-0.38}$}}	\\
        J060747.89+204120.2	&	G0	&	-12.4$\pm$2.6	&	1.1403	&	$1747.6${\raisebox{0.5ex}{\tiny$^{+1830.6}_{-1671.6}$}}	&	$0.2${\raisebox{0.5ex}{\tiny$^{+0.25}_{-0.2}$}}	&	$0.7${\raisebox{0.5ex}{\tiny$^{+0.61}_{-0.69}$}}	&	$-7.17${\raisebox{0.5ex}{\tiny$^{+0.54}_{-0.57}$}}	\\
        J055742.67+271929.1	&	G0	&	-12.8$\pm$1.9	&	1.0145	&	$1611.5${\raisebox{0.5ex}{\tiny$^{+1698.0}_{-1538.5}$}}	&	$0.4${\raisebox{0.5ex}{\tiny$^{+0.45}_{-0.35}$}}	&	$1.1${\raisebox{0.5ex}{\tiny$^{+1.1}_{-1}$}}	&	$-7.96${\raisebox{0.5ex}{\tiny$^{+0.56}_{-0.56}$}}	\\
        
    \end{tabular}
\end{table*}

\section{Summary}

In this work, we identified a sample of 260 CTTS from LAMOST DR6, most of which is spread towards the Galactic anti-center direction. The initial sample of 77,586 LELS was obtained from Edwin et al. (2022; under prep), based on the presence of H$\alpha$ emission line. To remove the giant stars from our sample, we used the log \textit{g} threshold of 3.5. Then, to identify young stars with IR excess, we made use of 2MASS-WISE CCDm after the prescribed quality cuts. Finally, we identified 260 bonafide CTTS from the initial sample, in which 104 stars are newly identified. It is seen that our sample of CTTS are spatially found to be within $100\degree \leq l \leq 220\degree$ and latitudes $-35\degree \leq |b| \leq15\degree$, which is also color-coded with respect to the star forming region associated with it. We used Gaia DR3 photometric values to estimate the stellar parameters such as mass and age of identified CTTS. The age range of CTTS is from 0.1 -- 20 Myr, with 90\% of the stars having age below 10 Myr. Also, the estimated mass range of the CTTS is between 0.11 -- 1.9 M\textsubscript{\(\odot\)}. Using the age and mass estimates, we derived the $\dot{M}_{acc}$ for 172 CTTS, which range from $10^{-7}$ -- $10^{-10}$ $M_{\odot}yr^{-1}$. The values are tabulated in Table \ref{tab:T1}. We find a clear trend of mass accretion rates increasing as a function of stellar mass. Irrespective of the location of the stars, either in Galactic center or Galactic anti-center direction, the power-law relationship obtained for our stars, $\dot{M}_{acc}$ $\propto$ $M_{*} ^ {1.43 \pm 0.26}$ is found to be in agreement with the literature. 

\section*{Acknowledgements}
We thank Cysil Baby Tom for his valuable suggestion in this work. The authors are grateful to the Centre for Research, CHRIST (Deemed to be University), Bangalore for the research grant extended to carry out the present project (MRPDSC-1932). Also, we would like to thank the Science \& Engineering Research Board (SERB), a statutory body of Department of Science \& Technology (DST), Government of India, for funding our research under grant number CRG/2019/005380. This work has made use of data products from the Guo Shoujing Telescope (the Large Sky Area Multi-Object Fibre Spectroscopic Telescope, LAMOST), and data from the European Space Agency (ESA) mission Gaia (https://www.cosmos.esa.int/gaia), processed by the Gaia Data Processing and Analysis Consortium (DPAC, https://www.cosmos.esa.int/web/gaia/dpac/consortium). Funding for the DPAC has been provided by national institutions, in particular the institutions participating in the Gaia Multilateral Agreement. We thank the SIMBAD database and the online VizieR library service for helping us in literature survey and obtaining relevant data.

\section*{Data Availability}

The data underlying this article was accessed from the Large sky Area Multi-Object fibre Spectroscopic Telescope (LAMOST) data release 6 (http://dr6.lamost.org/). Full version of Table 1 will be made available online in a machine-readable format.

\bibliography{ref} 

\begin{thebibliography}{}
\expandafter\ifx\csname natexlab\endcsname\relax\def\natexlab#1{#1}\fi

\bibitem[{{Alcal{\'a}} {$et~al$.}(2014){Alcal{\'a}}, {Natta}, {Manara},
  {Spezzi}, {Stelzer}, {Frasca}, {Biazzo}, {Covino}, {Randich}, {Rigliaco},
  {Testi}, {Comer{\'o}n}, {Cupani}, \& {D'Elia}}]{2014A&A...561A...2A}
{Alcal{\'a}}, J.~M., {Natta}, A., {Manara}, C.~F., {$et~al$.} 2014, Astronomy
  \& Astrophysics, 561, A2

\bibitem[{{Alcal{\'a}} {$et~al$.}(2017){Alcal{\'a}}, {Manara}, {Natta},
  {Frasca}, {Testi}, {Nisini}, {Stelzer}, {Williams}, {Antoniucci}, {Biazzo},
  {Covino}, {Esposito}, {Getman}, \& {Rigliaco}}]{2017A&A...600A..20A}
{Alcal{\'a}}, J.~M., {Manara}, C.~F., {Natta}, A., {$et~al$.} 2017, Astronomy
  \& Astrophysics, 600, A20

\bibitem[{{Andre} {$et~al$.}(1993){Andre}, {Ward-Thompson}, \&
  {Barsony}}]{1993ApJ...406..122A}
{Andre}, P., {Ward-Thompson}, D., \& {Barsony}, M. 1993, Astrophysical Journal,
  406, 122

\bibitem[{{Anusha} {$et~al$.}(2021){Anusha}, {Mathew}, {Shridharan}, {Arun},
  {Nidhi}, {Banerjee}, {Kartha}, {Paul}, \&
  {Bhattacharyya}}]{2021MNRAS.501.5927A}
{Anusha}, R., {Mathew}, B., {Shridharan}, B., {$et~al$.} 2021, Monthly Notices
  of the Royal Astronomical Society, 501, 5927

\bibitem[{{Appenzeller} \& {Mundt}(1989)}]{1989A&ARv...1..291A}
{Appenzeller}, I., \& {Mundt}, R. 1989, The Astronomy and Astrophysics Review,
  1, 291

\bibitem[{{Bailer-Jones} {$et~al$.}(2021){Bailer-Jones}, {Rybizki},
  {Fouesneau}, {Demleitner}, \& {Andrae}}]{2021AJ....161..147B}
{Bailer-Jones}, C.~A.~L., {Rybizki}, J., {Fouesneau}, M., {Demleitner}, M., \&
  {Andrae}, R. 2021, The Astronomical Journal, 161, 147

\bibitem[{{Barentsen} {$et~al$.}(2011){Barentsen}, {Vink}, {Drew}, {Greimel},
  {Wright}, {Drake}, {Martin}, {Valdivielso}, \&
  {Corradi}}]{2011MNRAS.415..103B}
{Barentsen}, G., {Vink}, J.~S., {Drew}, J.~E., {$et~al$.} 2011, Monthly Notices
  of the Royal Astronomical Society, 415, 103

\bibitem[{{Barrado y Navascu{\'e}s} \&
  {Mart{\'\i}n}(2003)}]{2003AJ....126.2997B}
{Barrado y Navascu{\'e}s}, D., \& {Mart{\'\i}n}, E.~L. 2003, The Astronomical
  Journal, 126, 2997

\bibitem[{{Bertout}(1989)}]{1989ARA&A..27..351B}
{Bertout}, C. 1989, Annual Rev. Astron. Astrophys, 27, 351

\bibitem[{{Bhattacharyya} {$et~al$.}(2022){Bhattacharyya}, {Mathew},
  {Ezhikode}, {Muneer}, {Selvakumar}, {Maheswer}, {Arun}, {Anilkumar},
  {Banerjee}, {Pramod}, {Kartha}, {Paul}, \& {Velu}}]{2022ApJ...933L..34B}
{Bhattacharyya}, S., {Mathew}, B., {Ezhikode}, S.~H., {$et~al$.} 2022, The
  Astrophysical Journal Letters, 933, L34

\bibitem[{{Bressan} {$et~al$.}(2012){Bressan}, {Marigo}, {Girardi},
  {Salasnich}, {Dal Cero}, {Rubele}, \& {Nanni}}]{2012MNRAS.427..127B}
{Bressan}, A., {Marigo}, P., {Girardi}, L., {$et~al$.} 2012, Monthly Notices of
  the Royal Astronomical Society, 427, 127

\bibitem[{{Briceno} {$et~al$.}(2019){Briceno}, {Calvet}, {Hernandez}, {Vivas},
  {Mateu}, {Downes}, {Loerincs}, {Perez-Blanco}, {Berlind}, {Espaillat},
  {Allen}, {Hartmann}, {Mateo}, \& {Bailey}}]{2019AJ....157...85B}
{Briceno}, C., {Calvet}, N., {Hernandez}, J., {$et~al$.} 2019, The Astronomical
  Journal, 157, 85

\bibitem[{{Casey} {$et~al$.}(2016){Casey}, {Ruchti}, {Masseron}, {Randich},
  {Gilmore}, {Lind}, {Kennedy}, {Koposov}, {Hourihane}, {Franciosini}, {Lewis},
  {Magrini}, {Morbidelli}, {Sacco}, {Worley}, {Feltzing}, {Jeffries},
  {Vallenari}, {Bensby}, {Bragaglia}, {Flaccomio}, {Francois}, {Korn},
  {Lanzafame}, {Pancino}, {Recio-Blanco}, {Smiljanic}, {Carraro}, {Costado},
  {Damiani}, {Donati}, {Frasca}, {Jofr{\'e}}, {Lardo}, {de Laverny}, {Monaco},
  {Prisinzano}, {Sbordone}, {Sousa}, {Tautvai{\v{s}}ien{\.{e}}}, {Zaggia},
  {Zwitter}, {Delgado Mena}, {Chorniy}, {Martell}, {Silva Aguirre}, {Miglio},
  {Chiappini}, {Montalban}, {Morel}, \& {Valentini}}]{2016MNRAS.461.3336C}
{Casey}, A.~R., {Ruchti}, G., {Masseron}, T., {$et~al$.} 2016, Monthly Notices
  of the Royal Astronomical Society, 461, 3336

\bibitem[{{Cieza} {$et~al$.}(2007){Cieza}, {Padgett}, {Stapelfeldt},
  {Augereau}, {Harvey}, {Evans}, {Mer{\'\i}n}, {Koerner}, {Sargent}, {van
  Dishoeck}, {Allen}, {Blake}, {Brooke}, {Chapman}, {Huard}, {Lai}, {Mundy},
  {Myers}, {Spiesman}, \& {Wahhaj}}]{2007ApJ...667..308C}
{Cieza}, L., {Padgett}, D.~L., {Stapelfeldt}, K.~R., {$et~al$.} 2007, The
  Astrophysical Journal, 667, 308

\bibitem[{{Cohen} \& {Kuhi}(1979)}]{1979ApJS...41..743C}
{Cohen}, M., \& {Kuhi}, L.~V. 1979, Astrophysical Journal, Suppl., 41, 743

\bibitem[{{Cottle} {$et~al$.}(2018){Cottle}, {Covey}, {Su{\'a}rez},
  {Rom{\'a}n-Z{\'u}{\~n}iga}, {Schlafly}, {Downes}, {Ybarra}, {Hernandez},
  {Stassun}, {Stringfellow}, {Getman}, {Feigelson}, {Borissova}, {Kim},
  {Roman-Lopes}, {Da Rio}, {De Lee}, {Frinchaboy}, {Kounkel}, {Majewski},
  {Mennickent}, {Nidever}, {Nitschelm}, {Pan}, {Shetrone}, {Zasowski},
  {Chambers}, {Magnier}, \& {Valenti}}]{2018ApJS..236...27C}
{Cottle}, J.~N., {Covey}, K.~R., {Su{\'a}rez}, G., {$et~al$.} 2018,
  Astrophysical Journal Supplement, 236, 27

\bibitem[{{Cui} {$et~al$.}(2012){Cui}, {Zhao}, {Chu}, {Li}, {Li}, {Zhang},
  {Su}, {Yao}, {Wang}, {Xing}, {Li}, {Zhu}, {Wang}, {Gu}, {Luo}, {Xu}, {Zhang},
  {Liu}, {Zhang}, {Yang}, {Cao}, {Chen}, {Chen}, {Chen}, {Chen}, {Chu}, {Feng},
  {Gong}, {Hou}, {Hu}, {Hu}, {Hu}, {Jia}, {Jiang}, {Jiang}, {Jiang}, {Jin},
  {Li}, {Li}, {Li}, {Liu}, {Liu}, {Lu}, {Mao}, {Men}, {Qi}, {Qi}, {Shi},
  {Tang}, {Tao}, {Wang}, {Wang}, {Wang}, {Wang}, {Wang}, {Wang}, {Wang},
  {Wang}, {Wang}, {Wang}, {Wang}, {Wang}, {Xu}, {Xu}, {Yang}, {Yu}, {Yuan},
  {Yuan}, {Zhai}, {Zhang}, {Zhang}, {Zhang}, {Zhao}, {Zhou}, {Zhou}, {Zhu}, \&
  {Zou}}]{2012RAA....12.1197C}
{Cui}, X.-Q., {Zhao}, Y.-H., {Chu}, Y.-Q., {$et~al$.} 2012, Research in
  Astronomy and Astrophysics, 12, 1197

\bibitem[{{Cutri} {$et~al$.}(2003){Cutri}, {Skrutskie}, {van Dyk}, {Beichman},
  {Carpenter}, {Chester}, {Cambresy}, {Evans}, {Fowler}, {Gizis}, {Howard},
  {Huchra}, {Jarrett}, {Kopan}, {Kirkpatrick}, {Light}, {Marsh}, {McCallon},
  {Schneider}, {Stiening}, {Sykes}, {Weinberg}, {Wheaton}, {Wheelock}, \&
  {Zacarias}}]{2003yCat.2246....0C}
{Cutri}, R.~M., {Skrutskie}, M.~F., {van Dyk}, S., {$et~al$.} 2003, VizieR
  Online Data Catalog, II/246

\bibitem[{{Cutri} {$et~al$.}(2021){Cutri}, {Wright}, {Conrow}, {Fowler},
  {Eisenhardt}, {Grillmair}, {Kirkpatrick}, {Masci}, {McCallon}, {Wheelock},
  {Fajardo-Acosta}, {Yan}, {Benford}, {Harbut}, {Jarrett}, {Lake}, {Leisawitz},
  {Ressler}, {Stanford}, {Tsai}, {Liu}, {Helou}, {Mainzer}, {Gettngs},
  {Gonzalez}, {Hoffman}, {Marsh}, {Padgett}, {Skrutskie}, {Beck}, {Papin}, \&
  {Wittman}}]{2014yCat.2328....0C}
{Cutri}, R.~M., {Wright}, E.~L., {Conrow}, T., {$et~al$.} 2021, VizieR Online
  Data Catalog, II/328

\bibitem[{{Evans} {$et~al$.}(2009){Evans}, {Dunham}, {J{\o}rgensen}, {Enoch},
  {Mer{\'\i}n}, {van Dishoeck}, {Alcal{\'a}}, {Myers}, {Stapelfeldt}, {Huard},
  {Allen}, {Harvey}, {van Kempen}, {Blake}, {Koerner}, {Mundy}, {Padgett}, \&
  {Sargent}}]{2009ApJS..181..321E}
{Evans}, Neal~J., I., {Dunham}, M.~M., {J{\o}rgensen}, J.~K., {$et~al$.} 2009,
  Astrophysical Journal Supplement, 181, 321

\bibitem[{{Fang} {$et~al$.}(2009){Fang}, {van Boekel}, {Wang}, {Carmona},
  {Sicilia-Aguilar}, \& {Henning}}]{2009A&A...504..461F}
{Fang}, M., {van Boekel}, R., {Wang}, W., {$et~al$.} 2009, Astronomy and
  Astrophysics, 504, 461

\bibitem[{{Feigelson} {$et~al$.}(2003){Feigelson}, {Gaffney}, {Garmire},
  {Hillenbrand}, \& {Townsley}}]{2003ApJ...584..911F}
{Feigelson}, E.~D., {Gaffney}, James~A., I., {Garmire}, G., {Hillenbrand},
  L.~A., \& {Townsley}, L. 2003, Astrophysical Journal, 584, 911

\bibitem[{{Frasca} {$et~al$.}(2017){Frasca}, {Biazzo}, {Alcal{\'a}}, {Manara},
  {Stelzer}, {Covino}, \& {Antoniucci}}]{2017A&A...602A..33F}
{Frasca}, A., {Biazzo}, K., {Alcal{\'a}}, J.~M., {$et~al$.} 2017, Astronomy and
  Astrophysics, 602, A33

\bibitem[{{Frasca} {$et~al$.}(2022){Frasca}, {Molenda-{\.Z}akowicz},
  {Alonso-Santiago}, {Catanzaro}, {De Cat}, {Fu}, {Zong}, {Wang}, {Cang}, \&
  {Wang}}]{2022A&A...664A..78F}
{Frasca}, A., {Molenda-{\.Z}akowicz}, J., {Alonso-Santiago}, J., {$et~al$.}
  2022, Astronomy and Astrophysics, 664, A78

\bibitem[{{Furlan} {$et~al$.}(2011){Furlan}, {Luhman}, {Espaillat},
  {D'Alessio}, {Adame}, {Manoj}, {Kim}, {Watson}, {Forrest}, {McClure},
  {Calvet}, {Sargent}, {Green}, \& {Fischer}}]{2011ApJS..195....3F}
{Furlan}, E., {Luhman}, K.~L., {Espaillat}, C., {$et~al$.} 2011, Astrophysical
  Journal, Suppl., 195, 3

\bibitem[{{Gaia Collaboration}(2022)}]{2022yCat.1355....0G}
{Gaia Collaboration}. 2022, VizieR Online Data Catalog, I/355

\bibitem[{{Green} {$et~al$.}(2019){Green}, {Schlafly}, {Zucker}, {Speagle}, \&
  {Finkbeiner}}]{2019ApJ...887...93G}
{Green}, G.~M., {Schlafly}, E., {Zucker}, C., {Speagle}, J.~S., \&
  {Finkbeiner}, D. 2019, The Astrophysical Journal, 887, 93

\bibitem[{{Guenther} {$et~al$.}(1999){Guenther}, {Lehmann}, {Emerson}, \&
  {Staude}}]{1999A&A...341..768G}
{Guenther}, E.~W., {Lehmann}, H., {Emerson}, J.~P., \& {Staude}, J. 1999,
  Astronomy and Astrophysics, 341, 768

\bibitem[{{Gullbring} {$et~al$.}(1998){Gullbring}, {Hartmann}, {Brice{\~n}o},
  \& {Calvet}}]{1998ApJ...492..323G}
{Gullbring}, E., {Hartmann}, L., {Brice{\~n}o}, C., \& {Calvet}, N. 1998, The
  Astrophysical Journal, 492, 323

\bibitem[{{Gutermuth} {$et~al$.}(2009){Gutermuth}, {Megeath}, {Myers}, {Allen},
  {Pipher}, \& {Fazio}}]{2009ApJS..184...18G}
{Gutermuth}, R.~A., {Megeath}, S.~T., {Myers}, P.~C., {$et~al$.} 2009,
  Astrophysical Journal Supplement, 184, 18

\bibitem[{{Hartigan} {$et~al$.}(1995){Hartigan}, {Edwards}, \&
  {Ghandour}}]{1995ApJ...452..736H}
{Hartigan}, P., {Edwards}, S., \& {Ghandour}, L. 1995, The Astrophysical
  Journal, 452, 736

\bibitem[{{Hartmann} {$et~al$.}(2006){Hartmann}, {D'Alessio}, {Calvet}, \&
  {Muzerolle}}]{2006ApJ...648..484H}
{Hartmann}, L., {D'Alessio}, P., {Calvet}, N., \& {Muzerolle}, J. 2006, The
  Astrophysical Journal, 648, 484

\bibitem[{{Hartmann} {$et~al$.}(2016){Hartmann}, {Herczeg}, \&
  {Calvet}}]{2016ARA&A..54..135H}
{Hartmann}, L., {Herczeg}, G., \& {Calvet}, N. 2016, Annual Rev. Astron.
  Astrophys, 54, 135

\bibitem[{{Hartmann} {$et~al$.}(1994){Hartmann}, {Hewett}, \&
  {Calvet}}]{1994ApJ...426..669H}
{Hartmann}, L., {Hewett}, R., \& {Calvet}, N. 1994, The Astrophysical Journal,
  426, 669

\bibitem[{{Herbig}(1960)}]{1960ApJS....4..337H}
{Herbig}, G.~H. 1960, Astrophysical Journal Supplement, 4, 337

\bibitem[{{Herbst} \& {Shevchenko}(1999)}]{1999AJ....118.1043H}
{Herbst}, W., \& {Shevchenko}, V.~S. 1999, The Astronomical Journal, 118, 1043

\bibitem[{{Herczeg} \& {Hillenbrand}(2008)}]{2008ApJ...681..594H}
{Herczeg}, G.~J., \& {Hillenbrand}, L.~A. 2008, The Astrophysical Journal, 681,
  594

\bibitem[{{Herczeg} \& {Hillenbrand}(2014)}]{2014ApJ...786...97H}
---. 2014, The Astrophysical Journal, 786, 97

\bibitem[{{Hillenbrand}(1997)}]{1997AJ....113.1733H}
{Hillenbrand}, L.~A. 1997, The Astronomical Journal, 113, 1733

\bibitem[{{Hou} {$et~al$.}(2016){Hou}, {Luo}, {Hu}, {Yang}, {Du}, {Liu}, {Lee},
  {Lin}, {Wang}, {Zhang}, {Cao}, \& {Hou}}]{2016RAA....16..138H}
{Hou}, W., {Luo}, A.~L., {Hu}, J.-Y., {$et~al$.} 2016, Research in Astronomy
  and Astrophysics, 16, 138

\bibitem[{{Hsieh} \& {Lai}(2013)}]{2013ApJS..205....5H}
{Hsieh}, T.-H., \& {Lai}, S.-P. 2013, Astrophysical Journal Supplement, 205, 5

\bibitem[{{Hutchings}(1970)}]{1970MNRAS.147..161H}
{Hutchings}, J.~B. 1970, Monthly Notices of the Royal Astronomical Society,
  147, 161

\bibitem[{{Kalari}(2019)}]{2019MNRAS.484.5102K}
{Kalari}, V.~M. 2019, Monthly Notices of the Royal Astronomical Society, 484,
  5102

\bibitem[{{Kalari} {$et~al$.}(2015){Kalari}, {Vink}, {Drew}, {Barentsen},
  {Drake}, {Eisl{\"o}ffel}, {Mart{\'\i}n}, {Parker}, {Unruh}, {Walton}, \&
  {Wright}}]{2015MNRAS.453.1026K}
{Kalari}, V.~M., {Vink}, J.~S., {Drew}, J.~E., {$et~al$.} 2015, Monthly Notices
  of the Royal Astronomical Society, 453, 1026

\bibitem[{{Koenig} \& {Leisawitz}(2014)}]{2014ApJ...791..131K}
{Koenig}, X.~P., \& {Leisawitz}, D.~T. 2014, The Astrophysical Journal, 791,
  131

\bibitem[{{Koenigl}(1991)}]{1991ApJ...370L..39K}
{Koenigl}, A. 1991, Astrophysical Journal Letters, 370, L39

\bibitem[{{Kogure} \& {Leung}(2007)}]{2007ASSL..342.....K}
{Kogure}, T., \& {Leung}, K.-C. 2007, {The Astrophysics of Emission-Line
  Stars}, Vol. 342, doi:10.1007/978-0-387-68995-1

\bibitem[{{Lada}(1987)}]{1987IAUS..115....1L}
{Lada}, C.~J. 1987, in Star Forming Regions, ed. M.~{Peimbert} \& J.~{Jugaku},
  Vol. 115, 1

\bibitem[{{Lee} {$et~al$.}(2005){Lee}, {Chen}, {Zhang}, \&
  {Hu}}]{2005ApJ...624..808L}
{Lee}, H.-T., {Chen}, W.~P., {Zhang}, Z.-W., \& {Hu}, J.-Y. 2005, The
  Astrophysical Journal, 624, 808

\bibitem[{{Liu} {$et~al$.}(2014){Liu}, {Yuan}, {Huo}, {Deng}, {Hou}, {Zhao},
  {Zhao}, {Shi}, {Luo}, {Xiang}, {Zhang}, {Huang}, \&
  {Zhang}}]{2014IAUS..298..310L}
{Liu}, X.~W., {Yuan}, H.~B., {Huo}, Z.~Y., {$et~al$.} 2014, in Setting the
  scene for Gaia and LAMOST, ed. S.~{Feltzing}, G.~{Zhao}, N.~A. {Walton}, \&
  P.~{Whitelock}, Vol. 298, 310--321

\bibitem[{{Lu} {$et~al$.}(2018){Lu}, {Zhang}, {Han}, \&
  {Shi}}]{2018Ap&SS.363..104L}
{Lu}, H.-p., {Zhang}, L.-y., {Han}, X.~L., \& {Shi}, J. 2018, Astrophysics and
  Space Science, 363, 104

\bibitem[{{Luo} {$et~al$.}(2015){Luo}, {Zhao}, {Zhao}, {Deng}, {Liu}, {Jing},
  {Wang}, {Zhang}, {Shi}, {Cui}, {Chu}, {Li}, {Bai}, {Wu}, {Cai}, {Cao}, {Cao},
  {Carlin}, {Chen}, {Chen}, {Chen}, {Chen}, {Chen}, {Chen}, {Chen},
  {Christlieb}, {Chu}, {Cui}, {Dong}, {Du}, {Fan}, {Feng}, {Fu}, {Gao}, {Gong},
  {Gu}, {Guo}, {Han}, {He}, {Hou}, {Hou}, {Hou}, {Hu}, {Hu}, {Hu}, {Huo},
  {Jia}, {Jiang}, {Jiang}, {Jiang}, {Jin}, {Kong}, {Kong}, {Lei}, {Li}, {Li},
  {Li}, {Li}, {Li}, {Li}, {Li}, {Li}, {Li}, {Li}, {Li}, {Li}, {Liang}, {Lin},
  {Liu}, {Liu}, {Liu}, {Liu}, {Lu}, {Luo}, {Mao}, {Newberg}, {Ni}, {Qi}, {Qi},
  {Shen}, {Shi}, {Song}, {Song}, {Su}, {Su}, {Tang}, {Tao}, {Tian}, {Wang},
  {Wang}, {Wang}, {Wang}, {Wang}, {Wang}, {Wang}, {Wang}, {Wang}, {Wang},
  {Wang}, {Wang}, {Wang}, {Wang}, {Wang}, {Wang}, {Wang}, {Wang}, {Wang},
  {Wang}, {Wei}, {Wei}, {Wu}, {Wu}, {Wu}, {Wu}, {Xing}, {Xu}, {Xu}, {Xu},
  {Yan}, {Yang}, {Yang}, {Yang}, {Yang}, {Yao}, {Yu}, {Yuan}, {Yuan}, {Yuan},
  {Yuan}, {Zhai}, {Zhang}, {Zhang}, {Zhang}, {Zhang}, {Zhang}, {Zhang},
  {Zhang}, {Zhang}, {Zhao}, {Zhou}, {Zhou}, {Zhu}, {Zhu}, {Zou}, \&
  {Zuo}}]{2015RAA....15.1095L}
{Luo}, A.~L., {Zhao}, Y.-H., {Zhao}, G., {$et~al$.} 2015, Research in Astronomy
  and Astrophysics, 15, 1095

\bibitem[{{Manara} {$et~al$.}(2016){Manara}, {Rosotti}, {Testi}, {Natta},
  {Alcal{\'a}}, {Williams}, {Ansdell}, {Miotello}, {van der Marel}, {Tazzari},
  {Carpenter}, {Guidi}, {Mathews}, {Oliveira}, {Prusti}, \& {van
  Dishoeck}}]{2016A&A...591L...3M}
{Manara}, C.~F., {Rosotti}, G., {Testi}, L., {$et~al$.} 2016, Astronomy and
  Astrophysics, 591, L3

\bibitem[{{Manara} {$et~al$.}(2017){Manara}, {Testi}, {Herczeg}, {Pascucci},
  {Alcal{\'a}}, {Natta}, {Antoniucci}, {Fedele}, {Mulders}, {Henning},
  {Mohanty}, {Prusti}, \& {Rigliaco}}]{2017A&A...604A.127M}
{Manara}, C.~F., {Testi}, L., {Herczeg}, G.~J., {$et~al$.} 2017, Astronomy and
  Astrophysics, 604, A127

\bibitem[{{Martell} {$et~al$.}(2021){Martell}, {Simpson}, {Balasubramaniam},
  {Buder}, {Sharma}, {Hon}, {Stello}, {Ting}, {Asplund}, {Bland-Hawthorn}, {De
  Silva}, {Freeman}, {Hayden}, {Kos}, {Lewis}, {Lind}, {Zucker}, {Zwitter},
  {Campbell}, {{\v{C}}otar}, {Horner}, {Montet}, \&
  {Wittenmyer}}]{2021MNRAS.505.5340M}
{Martell}, S.~L., {Simpson}, J.~D., {Balasubramaniam}, A.~G., {$et~al$.} 2021,
  MNRAS, 505, 5340

\bibitem[{{Mart{\'\i}n}(1998)}]{1998AJ....115..351M}
{Mart{\'\i}n}, E.~L. 1998, The Astronomical Journal, 115, 351

\bibitem[{{Mathew} {$et~al$.}(2018){Mathew}, {Manoj}, {Narang}, {Banerjee},
  {Nayak}, {Muneer}, {Vig}, {Pramod Kumar}, {Paul}, \&
  {Maheswar}}]{2018ApJ...857...30M}
{Mathew}, B., {Manoj}, P., {Narang}, M., {$et~al$.} 2018, The Astrophysical
  Journal, 857, 30

\bibitem[{{Muzerolle} {$et~al$.}(2003){Muzerolle}, {Hillenbrand}, {Calvet},
  {Brice{\~n}o}, \& {Hartmann}}]{2003ApJ...592..266M}
{Muzerolle}, J., {Hillenbrand}, L., {Calvet}, N., {Brice{\~n}o}, C., \&
  {Hartmann}, L. 2003, The Astrophysical Journal, 592, 266

\bibitem[{{Natta} {$et~al$.}(2006){Natta}, {Testi}, \&
  {Randich}}]{2006A&A...452..245N}
{Natta}, A., {Testi}, L., \& {Randich}, S. 2006, Astronomy and Astrophysics,
  452, 245

\bibitem[{{Pandey} {$et~al$.}(2013){Pandey}, {Eswaraiah}, {Sharma}, {Samal},
  {Chauhan}, {Chen}, {Jose}, {Ojha}, {Kesh Yadav}, \&
  {Chandola}}]{2013ApJ...764..172P}
{Pandey}, A.~K., {Eswaraiah}, C., {Sharma}, S., {$et~al$.} 2013, The
  Astrophysical Journal, 764, 172

\bibitem[{{Pecaut} \& {Mamajek}(2013)}]{2013ApJS..208....9P}
{Pecaut}, M.~J., \& {Mamajek}, E.~E. 2013, Astrophysical Journal, Suppl., 208,
  9

\bibitem[{{Rigliaco} {$et~al$.}(2012){Rigliaco}, {Natta}, {Testi}, {Randich},
  {Alcal{\`a}}, {Covino}, \& {Stelzer}}]{2012A&A...548A..56R}
{Rigliaco}, E., {Natta}, A., {Testi}, L., {$et~al$.} 2012, Astronomy and
  Astrophysics, 548, A56

\bibitem[{{Rosendhal}(1973)}]{1973ApJ...186..909R}
{Rosendhal}, J.~D. 1973, The Astrophysical Journal, 186, 909

\bibitem[{{Shridharan} {$et~al$.}(2021){Shridharan}, {Mathew}, {Nidhi},
  {Anusha}, {Arun}, {Kartha}, \& {Kumar}}]{2021RAA....21..288S}
{Shridharan}, B., {Mathew}, B., {Nidhi}, S., {$et~al$.} 2021, Research in
  Astronomy and Astrophysics, 21, 288

\bibitem[{{Stacy} {$et~al$.}(2009){Stacy}, {Greif}, \&
  {Bromm}}]{2009ASPC..419..339S}
{Stacy}, A., {Greif}, T.~H., \& {Bromm}, V. 2009, in Astronomical Society of
  the Pacific Conference Series, Vol. 419, Galaxy Evolution: Emerging Insights
  and Future Challenges, ed. S.~{Jogee}, I.~{Marinova}, L.~{Hao}, \& G.~A.
  {Blanc}, 339

\bibitem[{{Tody}(1986)}]{1986SPIE..627..733T}
{Tody}, D. 1986, in Society of Photo-Optical Instrumentation Engineers (SPIE)
  Conference Series, Vol. 627, Instrumentation in astronomy VI, ed. D.~L.
  {Crawford}, 733

\bibitem[{{{\v{S}}koda} {$et~al$.}(2020){{\v{S}}koda}, {Podsztavek}, \&
  {Tvrd{\'\i}k}}]{2020A&A...643A.122S}
{{\v{S}}koda}, P., {Podsztavek}, O., \& {Tvrd{\'\i}k}, P. 2020, Astronomy and
  Astrophysics, 643, A122

\bibitem[{{Weymann}(1963)}]{1963ARA&A...1...97W}
{Weymann}, R. 1963, Annual Rev. Astron. Astrophys, 1, 97

\bibitem[{{White} \& {Basri}(2003)}]{2003ApJ...582.1109W}
{White}, R.~J., \& {Basri}, G. 2003, The Astrophysical Journal, 582, 1109

\bibitem[{{Wu} {$et~al$.}(2011){Wu}, {Luo}, {Li}, {Shi}, {Prugniel}, {Liang},
  {Zhao}, {Zhang}, {Bai}, {Wei}, {Dong}, {Zhang}, \&
  {Chen}}]{2011RAA....11..924W}
{Wu}, Y., {Luo}, A.~L., {Li}, H.-N., {$et~al$.} 2011, Research in Astronomy and
  Astrophysics, 11, 924

\bibitem[{{Yoshida} {$et~al$.}(2008){Yoshida}, {Omukai}, \&
  {Hernquist}}]{2008Sci...321..669Y}
{Yoshida}, N., {Omukai}, K., \& {Hernquist}, L. 2008, Science, 321, 669

\bibitem[{{Zhang} {$et~al$.}(2022){Zhang}, {Luo}, {Jiang}, {Hou}, {Zuo}, {Du},
  {Li}, \& {Zhao}}]{2022ApJ...936..151Z}
{Zhang}, Y.-J., {Luo}, A.~L., {Jiang}, B., {$et~al$.} 2022, The Astrophysical
  Journal, 936, 151

\bibitem[{{Zhao} {$et~al$.}(2012){Zhao}, {Zhao}, {Chu}, {Jing}, \&
  {Deng}}]{2012arXiv1206.3569Z}
{Zhao}, G., {Zhao}, Y., {Chu}, Y., {Jing}, Y., \& {Deng}, L. 2012, arXiv
  e-prints, arXiv:1206.3569

\bibitem[{{Zhong} {$et~al$.}(2020){Zhong}, {Chen}, {Wu}, {Li}, {Bai}, \&
  {Hou}}]{2020A&A...640A.127Z}
{Zhong}, J., {Chen}, L., {Wu}, D., {$et~al$.} 2020, Astronomy and Astrophysics,
  640, A127

\bibitem[{{Zuckerman} {$et~al$.}(2014){Zuckerman}, {Vican}, \&
  {Rodriguez}}]{2014ApJ...788..102Z}
{Zuckerman}, B., {Vican}, L., \& {Rodriguez}, D.~R. 2014, The Astrophysical
  Journal, 788, 102

\end{thebibliography}
\newpage

\end{document}